\documentclass[11pt]{article}
\usepackage{acl}
\usepackage{times}
\usepackage{float}
\usepackage{booktabs}
\usepackage{array}
\usepackage[table]{xcolor}

\usepackage{latexsym}
\usepackage{todonotes}
\usepackage{amsthm}
\usepackage[most]{tcolorbox}
\usepackage[T1]{fontenc}

\usepackage[utf8]{inputenc}

\usepackage{microtype}
\usepackage{tikz}
\usepackage{inconsolata}

\usepackage{graphicx}
\usepackage{xcolor} 

\title{A Survey of LLM-Based Applications in Programming Education: Balancing Automation and Human Oversight}

\author{\hspace{-1cm} Griffin Pitts$^*$ \and  Anurata Prabha Hridi$^*$ \\\hspace{-1cm}North Carolina State University\\
  \hspace{-1cm}\texttt{\{wgpitts,aphridi\}@nscu.edu}\\ \And \hspace{0.8cm}
  Arun-Balajiee Lekshmi-Narayanan \\
  \hspace{0.8cm} University of Pittsburgh \\
  \hspace{0.8cm}\texttt{arl122@pitt.edu} \\
}
  
\begin{document}
\maketitle
\begingroup
\renewcommand\thefootnote{}\footnotetext{$^*$Both authors contributed equally to this work.}
\endgroup
\begin{abstract}
Novice programmers benefit from timely, personalized support that addresses individual learning gaps, yet the availability of instructors and teaching assistants is inherently limited. Large language models (LLMs) present opportunities to scale such support, though their effectiveness depends on how well technical capabilities are aligned with pedagogical goals. This survey synthesizes recent work on LLM applications in programming education across three focal areas: formative code feedback, assessment, and knowledge modeling. We identify recurring design patterns in how these tools are applied and find that interventions are most effective when educator expertise complements model output through human-in-the-loop oversight, scaffolding, and evaluation. Fully automated approaches are often constrained in capturing the pedagogical nuances of programming education, although human-in-the-loop designs and course-specific adaptation offer promising directions for future improvement. Future research should focus on improving transparency, strengthening alignment with pedagogy, and developing systems that flexibly adapt to the needs of varied learning contexts.

\end{abstract}

\section{Introduction}
Introductory programming courses serve as critical gateways to computer science and related STEM careers~\cite{whalley2020mathematics}, yet they present unique pedagogical challenges that contribute to high attrition rates \cite{petersen2016revisiting}. Students must simultaneously master syntax, develop computational thinking skills, and learn to debug complex logical errors, creating cognitive demands that often overwhelm novices. Success in these courses often depends on access to timely, targeted interventions, e.g., feedback, explanations, and guided problem-solving, that address individual learning gaps~\cite{marwan2020immediate, messer2023machine}. These personalized interventions are especially important because students exhibit varied practice behaviors, depending on their inclination toward problem-solving and/or example exploration~\cite{poh2025example}. Traditionally, this kind of support has been provided through human instruction, with teaching assistants (TAs) guiding problem-solving strategies, offering debugging help during office hours, and providing detailed feedback on assignments~\cite{markel2021inside}. 
However, the scalability of this human-centered model is limited. Large enrollment courses, increasingly common in CS education, strain the capacity of instructional staff to provide individualized attention~\cite{ahmed_feedback_2025}. Additionally, students may delay seeking help when they anticipate long wait times for TA feedback~\cite{gao2023too}, and TAs themselves face heavy workloads from simultaneous requests during office hours~\cite{gao2022you}.

Large language models (LLMs) have emerged as promising tools to automate aspects of programming education support by addressing the core challenges novice programmers face, including debugging code errors~\cite{lahtinen2005study}, repairing faulty code \cite{javier2021understanding}, obtaining timely feedback, and mastering foundational concepts to work with programming problems~\cite{ahmad2020programming, rivers2016learning}. Recent advances in natural language processing now make it possible to generate context-aware feedback, offer real-time debugging support, and adapt explanations to a student’s skill level~\cite{yousef2025begrading, zhong2024debug, chen2024teaching, lui2024gptutor}. Early work also shows the potential of Human--AI collaboration, where LLM outputs are refined or guided by educators to better align with pedagogical goals~\cite{hassany2024human}. Though promising, it is still unclear what best practices should guide the integration of LLMs into programming education and how responsibilities should be balanced between automation and human expertise.

The presented paper addresses this gap by reviewing recent applications of LLMs in programming education, focusing on how they are applied, the challenges that arise in practice, and the opportunities to align technical advances more closely with pedagogy.

\section{Methodology and Paper Selection}\label{sec:method}
To survey research on the use of LLMs in programming education, we reviewed publications across leading venues in computing education, HCI, and NLP, such as \texttt{SIGCSE, ITiCSE, ICER, LAK, EDM, CHI, EMNLP,} and related workshops. We focused on the 2021-2025 period, when LLM-driven systems first began to appear in educational settings.

From the surveyed literature, we identified three focal areas in which LLMs are being applied: formative code feedback, assessment, and knowledge modeling. Each aligns with a central pedagogical challenge in introductory programming courses, supporting student learning at scale in contexts where one-on-one guidance is difficult to provide. Within formative code feedback, we distinguish between approaches that generate hints and explanations to help students identify and understand their errors, and those that produce corrected code as examples or candidate repairs. Assessment focuses on grading and providing evaluative feedback at scale, while knowledge modeling seeks to represent what students know and how they progress in order to give instructors actionable insights through learning analytics. The distribution of reviewed papers across these areas is shown in Table~\ref{tab:paper_counts}. 

\begin{table}[H]
    \centering
    \begin{tabular}{|c|c|}
         \hline
         \textbf{Topic} & \textbf{Selected Papers} \\ \hline
         Formative Code Feedback & 20 \\ \hline
         Assessment & 14 \\ \hline
         Knowledge Modeling & 8 \\ \hline
    \end{tabular}
    \caption{Number of papers reviewed across three focal areas.}
    \label{tab:paper_counts}
\end{table}

\section{LLM Usage in Formative Code Feedback}\label{sec:feedback}

Debugging assistance is one of the most common reasons to seek help in programming courses, and improvements in code correctness after such support can substantially boost short-term performance~\cite{gao2022you}. However, the scalability limits of human-led help sessions have prompted researchers to explore how LLMs can extend this support in programming education. LLM-based feedback on student code is typically delivered in two complementary forms. In some cases, models generate explanations, hints, or scaffolding that help students locate and reason about their own programming mistakes. In others, models produce corrected code directly, offering candidate repairs or examples that students can study and compare against their own solutions. Both approaches aim to reduce the bottleneck of human-provided feedback, although they differ in the amount of agency they leave with the learner.

A large body of work has explored how LLMs can generate hints and explanations to guide student reasoning. Early evaluations of open-source models suggest that they can assist with syntax and minor semantic issues but continue to struggle with more complex bug localization and multi-line logic errors~\cite{majdoub2024debugging}. To address these challenges, researchers have developed systems that focus on scaffolding student reasoning. \textit{BugSpotter}, for instance, combines static analysis with LLM reasoning to create interactive debugging exercises for low-level programming languages~\cite{p2025bugspotter}. Iterative self-debugging loops have been proposed, where models run their generated code, collect execution feedback, and refine patches in multiple passes~\cite{chen2024teaching}. Adaptive scaffolding systems further extend these capabilities by monitoring learner progress and providing timely hints or explanations to break complex reasoning into manageable steps~\cite{oli2024towards}. Although such advances improve the accuracy and relevance of LLM-supported feedback, human oversight remains necessary to ensure that outputs align with pedagogical goals~\cite{ZUBAIR2025the}. For example, \textit{CodeAid}~\cite{kazemitabaar2024codeaid} found that while LLMs can accelerate support for students, direct answers without educator scaffolding risk undermining learning, highlighting the role of instructors in contextualizing automated feedback.

Another cluster of research investigates improving the clarity of error explanations to support student comprehension. Fine-tuned LLMs have demonstrated the ability to produce clearer, context-sensitive error messages, improving novice problem-solving~\cite{vassar2024fine, leinonen-LLM-ErrorMessages}. Comparative analyses of human and model debugging strategies reveal differences in reasoning patterns, pointing to opportunities for designing AI-assisted tools that nudge learners toward more expert-like approaches~\cite{macneil2024decoding}. Related work on prompt engineering and explainability techniques, such as step-by-step runtime verification, has also shown promise for improving the readability of error messages and fostering trust in human–AI collaboration~\cite{zhong2024debug, hoq2025facilitating, kang2025explainable}. Expanding further, researchers have also applied LLMs to code quality feedback, detecting issues such as misleading variable and function identifiers in novice code~\cite{vrechtavckova2025finding}.

Beyond generating hints and explanations, a growing body of work explores producing corrected code and worked examples that students can study alongside their own solutions. For instance, LLMs have been applied to generate code-tracing questions for introductory courses, producing diverse and pedagogically aligned items~\cite{fan2023exploring}. Recent work demonstrates how LLMs can generate worked examples that help students learn strategies and better understand solution approaches~\cite{sarsa2022automatic}. Research also shows that students learn from the process of spotting and fixing code errors~\cite{koutcheme2024using}, and that these skills strongly predict learning outcomes and course success~\cite{gao2022you, gao2023too}. 

To support this process, automated program repair (APR) systems have targeted syntactic and semantic errors in student submissions, with LLMs broadening the scope of repairs to be more context-sensitive and benchmarked transparently~\cite{jiang2023impact}. Examples include \textit{PyDex}~\cite{zhang2024pydex}, which generates accurate leverages LLMs to automatically generate accurate fixes for common novice errors in Python assignments~\cite{zhang2024pydex}; \textit{COAST}, a multi-agent framework that coordinates detection, repair, and verification while synthesizing debugging datasets~\cite{yang2025coast}; and \textit{RepairLLaMA}, which incorporates repair-aware representations and parameter-efficient fine-tuning to outperform vanilla prompting on standard APR benchmarks~\cite{silva2025repairllama}.

While these advances demonstrate the technical potential of LLMs for formative programming support, their educational value depends on when and how the feedback is delivered. Automated fixes that come too early, solve too much of the problem, or present complete answers can short-circuit the learning process by removing opportunities for students to reason through their own errors. In contrast, tools that generate hints, scaffold reasoning, and explain errors without directly supplying solutions are better aligned with pedagogical goals. The challenge is therefore not only improving accuracy on complex bugs but also designing systems that adapt the level of support to the learner’s needs. Emerging best practices point toward hybrid approaches, where LLMs address routine or surface-level issues and generate scalable practice materials, while human educators provide context, address deeper misconceptions, and guide students toward lasting debugging strategies.

\section{LLM Usage in Assessment}\label{sec:support}
With the increasing availability of LLMs in education, there are now provisions for the use of automated teaching assistants (TAs) in assessments, particularly in programming courses where grading is frequent and labor-intensive~\cite{mehta2023can}. Early evaluations benchmarked the ability of LLMs to provide such feedback, demonstrating that even in zero-shot configurations, they can produce rubric-aligned evaluations with moderate agreement to human graders~\cite{yeung2025zero, silva2025assessing}. These findings position LLMs as viable tools for scalable deployment, reducing the need for extensive rule-based assessment design. For example, \textit{ABScribe}~\cite{reza2024abscribe} demonstrates how LLMs can support human-AI co-writing by generating and organizing multiple text variations, easing TA workload and improving revision efficiency. 

However, meta-analytic perspectives caution that such systems inherit biases from training data and require prompt and rubric alignment to meet course-specific standards~\cite{messer2023machine}. In classroom settings, results have been mixed. In a study involving more than 1,000 students, GPT-4 reliably evaluated straightforward and clear-cut submissions but required human arbitration for nuanced cases~\cite{chiang2024large}. Similarly, automated grading with LLMs in a bioinformatics course reduced TA workload and accelerated grading speed, but raised concerns about transparency, reproducibility, and student trust in AI-generated assessments~\cite{polivcar2025automated}. To address this, frameworks like \textit{BeGrading}~\cite{yousef2025begrading} have integrated LLMs into a multi-stage feedback pipeline, combining initial automated grading with targeted suggestions for improvement, while \textit{CodEv}~\cite{tseng2024codev} applied chain-of-thought prompting, ensemble reasoning, and consistency checks to produce accurate and constructive feedback. Other work, such as the AI-augmented TA feasibility study~\cite{ahmed_feedback_2025}, examined how LLMs can fit into human TA workflows in CS1 courses, focusing on providing timely, individualized support while preserving grading quality.  

Beyond grading accuracy, the specificity and pedagogical usefulness of LLM-generated feedback vary considerably~\cite{pankiewicz2023large, estevez2024evaluation}. Recent studies have examined how models can generate formative, actionable feedback that supports skill development in introductory programming courses~\cite{mehta2023can}. One line of work uses program repair tasks as a proxy for feedback quality, showing that automated grading outputs can contribute to improvements in students’ problem-solving and code comprehension skills~\cite{pankiewicz2023large}. Others highlight the need for careful prompt engineering, rubric alignment, and iterative evaluation to ensure that feedback remains contextually relevant and educationally valuable. Therefore, human oversight remains a central design principle in this space, with LLMs serving as collaborators that augment TA capacity rather than replacements.

\section{LLM Usage in Knowledge Modeling}\label{sec:kc}
Just as LLMs have been applied to formative feedback and assessment, they are increasingly being explored for a broader challenge in programming education: modeling what students know and how their understanding develops over time. Knowledge modeling supports instructional design by making student learning more visible, helping educators monitor progress, identify misconceptions, and create targeted interventions at scale. One common approach is knowledge component (KC) extraction, where student work is mapped to the concepts they need to master, such as variables, loops, and conditionals. While this process helps educators monitor progress and create targeted interventions, performing it manually is time-consuming and limits scalability.

\begin{table*}[t]
    \centering
    \begin{tabular}{p{4.6cm}p{10.4cm}}
        \toprule
        \rowcolor{gray!15}
        \textbf{Topic} & \textbf{Best Practices} \\
        \midrule
        Formative Code Feedback & {Use LLMs to generate hints, explanations, and error messages that scaffold in pedagogically-sound ways; balance automation with formative scaffolding to prevent over-complete fixes and overreliance.} \\
        Assessment & {Ensure LLM grading aligns with rubrics and course standards; use human arbitration for nuanced cases.} \\
        Knowledge Modeling & {When using LLMs for KC extraction and clustering, validate outputs against expert review of topics and subtopics.} \\
        \bottomrule
    \end{tabular}
    \caption{Best practices for LLM applications in programming education, synthesized across three focal areas.}
    \label{tab:best_practices}
\end{table*}

Recent advances have demonstrated how LLMs can automate this extraction process with promising results. Researchers used GPT-4 to generate and tag KCs from multiple-choice questions, with human evaluators preferring the LLM-generated tags over instructor-assigned ones in about two-thirds of cases~\cite{moore2024automated}. \textit{KCluster}~\cite{wei2025kcluster} is another approach to combine LLM-generated question similarity metrics with clustering algorithms to automatically group related problems and discover their underlying KCs, producing models that outperform expert-designed baselines. Additionally, others~\cite{niousha2025llm, o2025code, mittal2025modeling} have presented early successful results on the use of LLMs toward KC extraction.

Researchers have also explored how LLMs can perform KC extraction during real-time learning interactions. LLMs can annotate student-tutor dialogues with KC tags during conversations, achieving close to human-level accuracy~\cite{scarlatos2025exploring}. To gain more granular insights into student understanding, test case-informed knowledge tracing is another approach where individual test case pass/fail results serve as indicators for LLMs to better distinguish which concepts students have mastered versus those they struggle with~\cite{duan2025test}. Additionally, incorporating student self-reflection prompts can significantly improve KC tagging performance by LLMs~\cite{li2024automate}. 

While these advances highlight the potential of LLMs to scale knowledge modeling, their value depends on expert validation of concept mappings and alignment with course objectives. If not carefully validated, knowledge models that are inaccurate or poorly contextualized can misguide instructors and weaken their ability to design effective interventions. When integrated responsibly, however, LLM-generated models can strengthen learning analytics by giving educators actionable insights into student progress, revealing common misconceptions, and informing the design of targeted supports at scale. Future research can focus on improving the accuracy and reliability of these approaches across varied datasets and on developing methods that ensure valid and useful representations of student knowledge.

\section{Discussion}
Our review of recent LLM applications in programming education indicates that systems that successfully address students' pedagogical needs tend to retain human involvement throughout the workflow. Across formative code feedback, assessment, and knowledge modeling, successful applications frequently incorporate educators in the workflow to interpret results, refine automated outputs, or make instructional decisions. In contrast, fully automated systems often focus on narrower or more technical tasks where less contextual judgment is required. Best practices are summarized in Table~\ref{tab:best_practices}.

Altogether, the systems surveyed demonstrate several shared strengths. They address scalability by automating tasks that would otherwise demand substantial instructor time, such as grading large cohorts or generating individualized debugging hints. Open-source language models have also been incorporated into APR pipelines, where evaluation frameworks use GPT-4-as-a-judge to approximate expert review at scale. This approach highlights the benefits of mixed human-and-automated evaluation in balancing accuracy, scalability, and cost in this domain~\cite{koutcheme2024open, Koutcheme-LLMs4Feedback}. Research also incorporates mechanisms that improve consistency and transparency in instructional support, as seen in \textit{BeGrading}’s variance analysis and criteria-aligned feedback generation~\cite{yousef2025begrading}. Increasingly, these tools integrate pedagogical considerations into their design, from scaffolding strategies in debugging systems to feedback phrased in ways that guide student reflection and self-correction.

However, at the same time, performance varies considerably depending on factors such as prompt design, availability of course-specific training data, and evaluation practices. Many LLM-based systems are not explicitly tuned to instructional objectives, which can lead to technically correct but educationally unhelpful feedback~\cite{sonkar2024pedagogical}. Performance often drops when moving from controlled benchmarks to authentic, noisy student code, and the opacity of model reasoning can reduce trust among both students and instructors. There is also the risk that students may overrely on incorrect model outputs, undermining opportunities for productive learning interactions~\cite{pitts2025students}. 

\section{Limitations and Future Work}
The current body of research on LLMs in programming education is still in its early stages, with limitations that can guide future research. While the focal areas covered in this survey reflect active areas of research, other domains, such as collaborative coding and accessibility support, remain underexplored. Addressing these gaps can take place in a larger-scale systematic literature review. Additionally, relating to the maturity of the current work reviewed, many of the systems surveyed are early-stage prototypes or evaluated only in small-scale settings, with limited evidence on scalability or long-term learning outcomes.

Technically, while early work has investigated fine-tuning and reinforcement learning with human feedback (RLHF)~\cite{hicke2023ai}, there remains significant scope for advancing model development and designing workflows explicitly aligned with pedagogical goals through course-specific fine-tuning. Research could also investigate adaptive collaboration frameworks where the degree of automation varies according to task complexity, user proficiency, and the model’s own confidence in its output. Further priorities include identifying and mitigating biases in model outputs, especially in grading and feedback, and expanding the use of multi-modal, context-aware interaction that can adapt feedback to the learner’s current state. While LLMs can offer an improved learning experience for programming education, their greatest potential lies in augmenting rather than replacing human expertise. Systems that remain adaptable, transparent, and closely aligned with pedagogical best practices are most likely to deliver meaningful and sustainable benefits for learners.

% \subsection{Limitations} In this work, we selected topics where LLMs could be used in computer science education since researchers have explored these topics extensively recently. However, this field is ever evolving, and the number of relevant topics will increase.  

\bibliography{references}

\end{document}